\documentclass[aps,showpacs]{revtex4-1}

\usepackage{units}
\usepackage{graphicx}
\usepackage{float} 
\usepackage{amsmath}   %<-- To use \text comand inside a math text 
\usepackage{upgreek}  %<-- Use \upmu to avoid capital italicized Greek letters
\usepackage{bm}  % <-- allows you to write bold greek letters
\usepackage[latin1]{inputenc}
\usepackage{epstopdf}

\begin{document}

\title{Measuring The Heat Capacity in a Bose-Einstein Condensation using Global Variables}

\author{R.~F.~Shiozaki$^1$}
\author{G.~D.~Telles$^1$}\email{gugs@ifsc.usp.br}
\author{P.~Castilho$^1$}
\author{F.~J.~Poveda-Cuevas$^1$}
\author{S.~R.~Muniz$^1$}
\author{G.~Roati$^2$}
\author{V.~Romero-Rochin$^3$}
\author{V.~S.~Bagnato$^1$}

\affiliation{$^1$Instituto de F\'{\i}sica de S\~{a}o Carlos, Universidade de S\~{a}o Paulo, C.P. 369, 13560-970 S\~{a}o Carlos, SP, Brazil \\ $^2$LENS and Dipartamento di Fisica, Universit\'a di Firenze, and INFM-CNR, Via Nello Carrara 1, 50019 Sesto Fiorentino, Italy \\ $^3$Instituto de F\'{\i}sica, Universidad Nacional Aut\'onoma de M\'exico, Apartado Postal 20-364, 01000 M\'exico, D.F., Mexico}

\begin{abstract}

Phase transitions are well understood and generally followed by the behavior of the associated thermodynamic quantities, such as in the case of the $\lambda$ point superfluid transition of liquid helium, which is observed in its heat capacity. In the case of a trapped Bose-Einstein condensate (BEC), the heat capacity cannot be directly measured. In this work, we present a technique able to determine the global heat capacity from the density distribution of a weakly interacting gas trapped in an inhomogeneous potential. This approach represents an alternative to models based on local density approximation. By defining a pair of global conjugate variables, we determine the total internal energy and its temperature derivative, the heat capacity. We then apply the technique to a trapped $^{87}$Rb BEC a $\lambda$-type transition dependent on the atom number is observed, and the deviations from the non-interacting, ideal gas case are discussed. Finally we discuss the chances of using this method to study the heat capacity at $T \rightarrow 0$.

\end{abstract}

\maketitle

\section{Introduction}

The heat capacity is one of the fundamental quantities that contain information concerning the nature of a phase transition. The remarkable discontinuity in the heat capacity near the superfluid transition ($\lambda$ point) observed in liquid He was considered as one of the most important properties to be well understood, and the quest to understand it brought theoretical proof that Bose-Einstein condensation~(BEC) does take place in a liquid such as the superfluid He, please see Refs.~\cite{London1950,Feynman1953a}, and the references therein.

Even before the experimental observation of a BEC in trapped dilute gases~\cite{Anderson1995,Bradley1995,Davis1995}, predictions of the heat capacity in such systems were made~\cite{Bagnato1987}, and soon after its realization a measurement was performed by means of the release energy~\cite{Ensher1996}. Many theoretical works have risen during the past decades on this topic~\cite{Grossmann1995,Haugerud1997,Napolitano1997,Giorgini1997,Courteille2001}, but few experimental explorations of the heat capacity of trapped gases were performed due to a lack of methods to measure the total internal energy of the sample. Recently, the interest in heat capacity measurements rose back, mainly motivated by characterization of quantum degenerate gases at the unitary limit, where a strong correlated system is studied~\cite{Ku2012}. The understanding of the behavior of a strong interacting ensemble of quantum particles is a challenge to modern physics~\cite{Griffin1993,Bloch2008}.

The majority of the reported heat capacity measurements of trapped quantum gases across the phase transition, have used models based on local density approximation~(LDA)~\cite{Nascimbene2010,Ku2012}, which handles the inhomogeneous number density distribution in an inexact way. In this report we introduce a technique to measure a global (rather than local) heat capacity at ``constant volume'', $C_{\mathcal{V}}$. We have applied a newly developed idea based on global thermodynamic variables~\cite{Romero-Rochin2005,Sandoval-Figueroa2008} instead of LDA. The technique was used to measure $C_{\mathcal{V}}$ across the BEC transition of a $^{87}$Rb BEC confined in a harmonic magnetic trap. We start by presenting the theoretical background of our method, followed by a short description of our experimental setup, and then the main results and a discussions, including the effect of interactions.

\section{Global Variable Analysis and Heat Capacity}

In a recent publication~\cite{Romero-Rochin2012}, we have studied the BEC transition in a harmonically trapped $^{87}$Rb sample in terms of the new global thermodynamic parameters introduced in \cite{Romero-Rochin2005,Sandoval-Figueroa2008}. We have built a phase diagram split in two domain regions: \textit{(i)} a pure thermal gas; and, \textit{(ii)} a mixture of condensate and thermal fractions. By considering a collection of global variables ($N,T,\cal V$), where ${\cal V}=1/\omega^3$ is defined as \textit{volume parameter}, and $\omega=(\omega_x \omega_y \omega_z)^{\frac{1}{3}}$ is the geometric mean of the trapping frequencies, we defined $\Pi(N,T,\cal V)$ as \textit{pressure parameter}, which corresponds to the hydrostatic pressure of the system. In fact, $\Pi$ is the conjugate variable to $\cal V$, i.e.
\begin{equation}\label{eq:Helm}
\Pi=-\left( \frac{\partial F}{\partial {\cal V}} \right)_{N,T}
\end{equation}
for the Helmholtz free energy $F=F(N,T,\cal V)$. In this context, $\Pi$ is obtained as~\cite{Romero-Rochin2005,Sandoval-Figueroa2008}
\begin{equation}\label{eq:piform}
\Pi = \frac{2}{3{\cal V}} \int n(\vec r) \> \frac{1}{2} m (\vec \omega \cdot \vec r)^2 \> d^3 r ,
\end{equation}
which can be determined by knowing the density distribution $n(\vec{r})$ and the confining harmonic frequencies $\omega=(\omega_x \omega_y \omega_z)^{\frac{1}{3}}$. It is important to stress that Eq.~\ref{eq:piform} is valid in the thermodynamic limit $N \rightarrow \infty$ and ${\cal V} \rightarrow \infty$ $(\omega \rightarrow 0)$, which means that \textit{(i)} finite number effects are not considered, \textit{(ii)} $\hbar \omega$ must be much smaller than any other energy scale, and \textit{(iii)} the s-wave scattering length must be much smaller than the simple harmonic oscillator length~(short-range interactions). Also it is worth noting that Eq.~\ref{eq:piform} is written for a harmonic trapping potential, but the method can be applied to an arbitrary external potential~\cite{Sandoval-Figueroa2008}.

As a function of the global extensive parameter $\cal V$ and its intensive conjugate $\Pi$, one can show that the internal energy of a harmonically trapped thermal cloud and pure BEC are, respectively~\cite{ShiozPhD}: $U_{th}=3\Pi\mathcal{V}$ and $U_0=\frac{5}{2}\Pi\mathcal{V}$. Therefore it is possible to separate the contributions of the thermal and condensate components to the total internal energy as
\begin{equation}\label{eq:internalenergy}
U=3\Pi_{th}\mathcal{V}+\frac{5}{2}\Pi_0\mathcal{V}=3\Pi\mathcal{V}-\frac{1}{2}\Pi_0\mathcal{V},
\end{equation}
where the total \textit{pressure parameter} is also considered as the sum of the two components, $\Pi=\Pi_{th}+\Pi_0$. Then the heat capacity at constant \textit{volume parameter} can be calculated as $C_{\cal V}=\left(\frac{\partial U}{\partial T}\right)_{N,{\cal V}}$. By assuming $\frac{\partial\Pi_0}{\partial T}\ll \frac{\partial\Pi}{\partial T}$, the last term in Eq.~\ref{eq:internalenergy} can be neglected. In fact, our experimental data shows that $\left| \frac{\partial\Pi_0}{\partial T}/ \frac{\partial\Pi}{\partial T} \right| < 0.1$ even at the lowest temperature values of $T/T_c \approx 0.1$. Therefore we finally find that a good approximation is:
\begin{equation}\label{eq:cv}
C_{\cal V}= \frac{3}{\omega^3} \left( \frac{\partial \Pi}{\partial T}\right)_{N,\omega},
\end{equation}
which means that the heat capacity can be directly obtained by measuring the equation of state $\Pi(N/{\cal V},T)$ as we performed in~\cite{Romero-Rochin2012}.

Soon after the production of the first experimental BEC by the JILA group, an attempt to map the heat capacity was carried out, taking into account the balance between the internal energy and the kinetic energy released during the free fall~\cite{Ensher1996}. In that paper, the overall scaled energy per particle is obtained as a function of the temperature. Above the critical temperature the linearity of the energy with temperature indicates the Maxwell-Boltzmann classical limit. The data around the critical temperature suggest a change in the energy balance, indicating the jump in the heat capacity. The value extracted from the data is smaller than that expected for an ideal gas, but in good agreement with that predicted by a finite number corrected, ideal gas theory~\cite{Grossmann1995}.

From the theoretical point of view, the heat capacity near the phase transition has been derived using different approaches~\cite{Grossmann1995,Haugerud1997,Napolitano1997,Giorgini1997} in the presence and/or the absence of interactions. For particles confined in a harmonic potential, those calculations presented similar results. In this work, we will use, as a reference, the original calculation presented by~\cite{Bagnato1987}. That is, a Bose gas with large number of particles and negligible interactions. In this picture, the heat capacity, $C_{\cal V}$, evolves presenting a $\lambda$-shaped curve across the BEC transition for a harmonically trapped Bose gas. It displays a steep change in the $C_{\cal V}$ values, near the critical temperature. By defining $C^{-}_{\cal V} \equiv C_{\cal V}(T_c^{-})$, $C^{+}_{\cal V} \equiv C_{\cal V}(T_c^{+})$, and $\Delta C_{\cal V}=C^{-}_{\cal V}-C^{+}_{\cal V}$, the $C_{\cal V}$ peak value is found just below the critical temperature, given by: $\frac{C^{-}_{\cal V}}{Nk_B}=12\frac{\zeta(4)}{\zeta(3)}(\approx 10.8)$, and shall quickly change around $T_c$ by $\frac{\Delta C_{\cal V}}{Nk_B}=9\frac{\zeta(3)}{\zeta(2)}(\approx 6.6)$~\cite{Bagnato1987,Grossmann1995}. These theoretical results are general and valid for arbitrary oscillator frequencies.  The presence of weak interactions would produce minor changes on $C_{\cal V}$ very near the critical temperature when compared to the ideal Bose gas case, while keeping its overall shape~\cite{Ensher1996,Giorgini1997,Courteille2001}.

The resemblance to the liquid $^{4}He$ heat capacity, near the $\lambda$-point, is clear~\cite{Arp2005}, as well as its discontinuous jump on $T_{c}$, in the limit of very large $N$. This qualitative characteristic was predicted by different theories whether considering finite number effects~\cite{Grossmann1995,Napolitano1997} and interactions~\cite{Giorgini1997} or not~\cite{Bagnato1987}. However, for an ideal Bose gas in the large number limit, the values of $C^{+}_{\cal V}$, $C^{-}_{\cal V}$ and $\Delta C_{\cal V}$ scale linearly with the number of atoms. In fact, the interactions are predicted to  change the $C_{\cal V}$ behavior by rounding off the peak existing just below the critical temperature, as discussed by Giorgini~\cite{Giorgini1997}.

\section{Experimental Description, Results and Discussion}

The experimental system comprises a double magneto--optical trap (MOT) and a QUIC magnetic trap. Details about the system are described in previous publications~\cite{Henn2008}. A combination of laser cooling and RF-evaporative cooling allows to obtain a BEC containing $2-8 \times 10^5$ $^{87}$Rb atoms. The trap frequencies are $\omega_x=2\pi \times 23$Hz for the weak axis, and $\omega_y=\omega_z=2\pi \times 209$Hz for the most confining directions. A full characterization of the condensate is performed by, first, recording an absorption image with a CCD camera, after $15$ms of free expansion. The data was fitted using a bimodal atomic distribution (condensate and thermal fractions as a sum of a Thomas-Fermi and a Gaussian profiles, respectively), and we assume cylindrical symmetry for the trapping potential in order to rebuild the complete 3D distribution, i.e. the Gaussian widths and the Thomas-Fermi radii, as well as the respective error values, which account for deviations from both the ideal Gaussian assumption and the Thomas-Fermi approximation. 

\begin{figure}
\centering
 \includegraphics[width=1.0\textwidth]{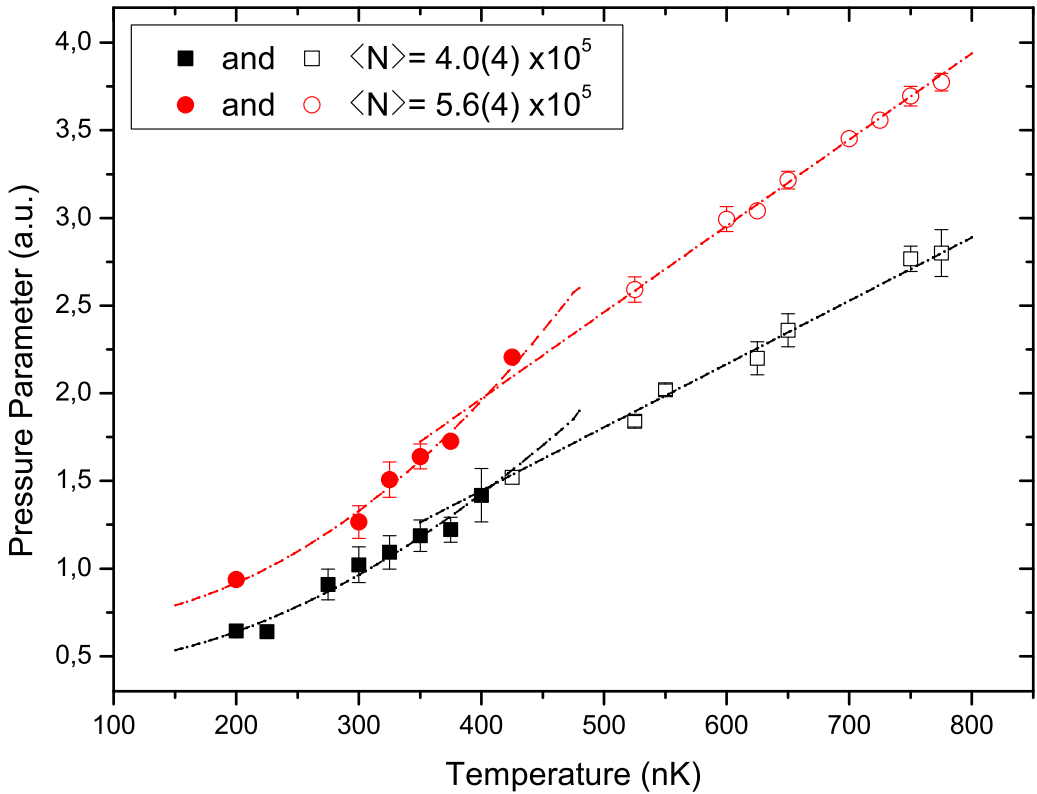}
\caption{
\textit{Pressure parameter} as a function of temperature for two different averaged particle number, $\langle N \rangle$. For each $\langle N \rangle$, the crossing between a linear fitting for thermal points (open symbols) and the interpolation/extrapolation of points with measurable condensate fraction (solid symbols) determines the critical point.}
\label{fig:pvst}
\end{figure}

The fittings also promptly provide the temperature values and both the condensate and thermal number of atoms. Then the \textit{in situ} distribution is determined by assuming a ballistic expansion of the thermal cloud, and then rescaling backwards the characteristic parabolic shape of the BEC expansion, as demonstrated in \cite{Castin1996,Kagan1996,Dalfovo1997}. For the BEC fraction, simple analytical expressions are known for an elongated cigar-shaped trap~\cite{Castin1996}. Finally the full 3D density distribution $n(\vec{r})$ is inserted in Eq.~\ref{eq:piform}, and the parameter $\Pi$ is obtained. Finite temperature corrections to the Thomas-Fermi (TF) expansion fits of our samples were neglected. We estimate these corrections would slightly change the measured radii, on the order of ~1\% or less.

\begin{figure}
\centering
 \includegraphics[width=0.9\textwidth]{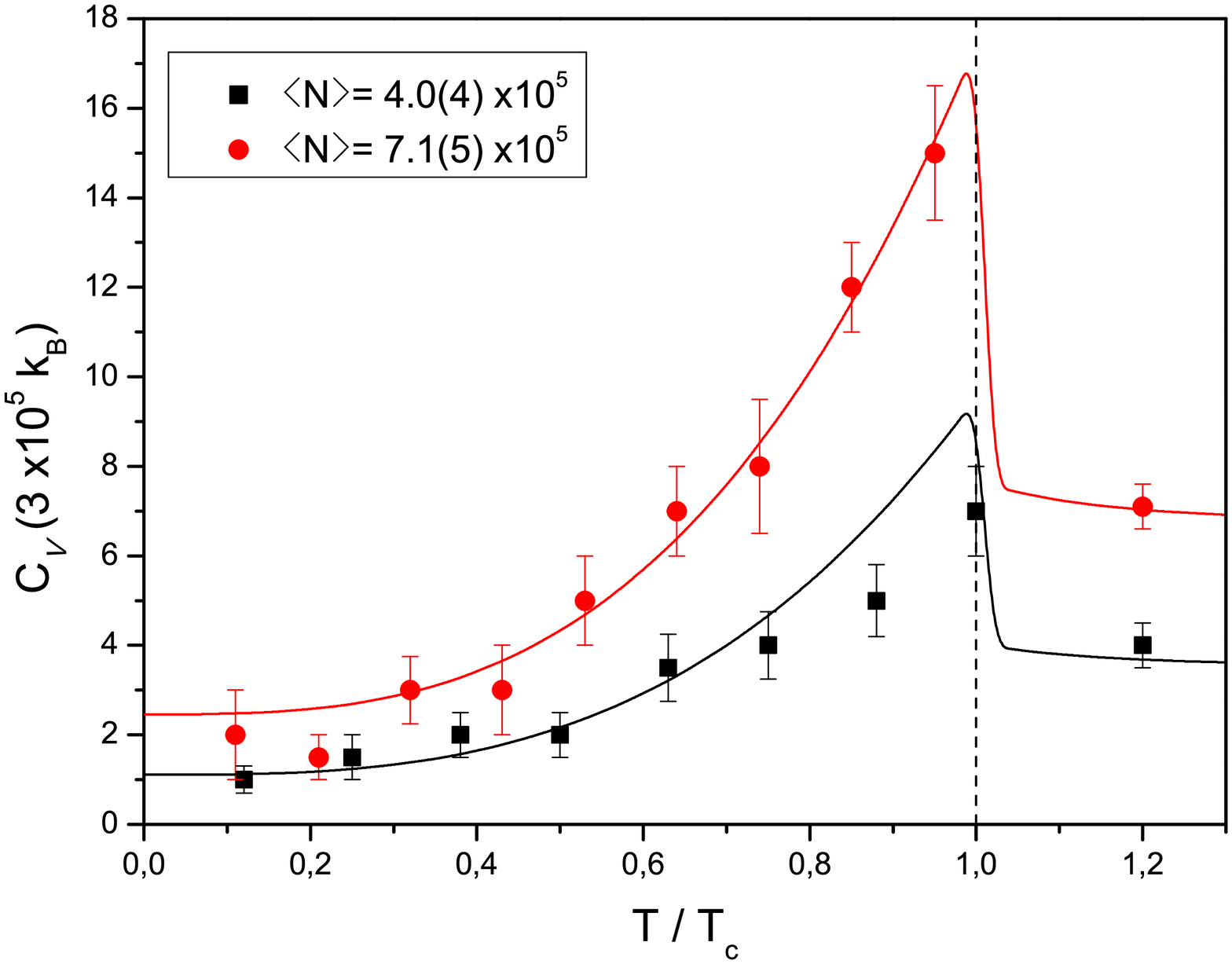}
\caption{
The heat capacity versus the normalized temperature is plotted around the condensation temperature, $T_{c}$, for two different atom numbers. The experimental points are extracted from diagram of Fig.~\ref{fig:pvst}, via Eq.~\ref{eq:cv}. The lines result from Eqs.~10 and 14 ~\cite{Grossmann1995}, where $N$ is the only adjustable parameter}
\label{fig:cv}
\end{figure}

The heat capacity is determined as follows. First, we derive the equation of state $\Pi=\Pi(N/{\cal V},T)$ by taking data in a constant volume trap potential ($\cal V$ constant). The acquired data sets taken for the same number of atoms, $N$, are grouped. We then plotted isodensity curves, $\Pi$ versus $T$, for different atom numbers, $\langle N \rangle$, as shown in Fig.~\ref{fig:pvst}. From those $\Pi$ versus $T$ graphs, we finaly determine the heat capacity $C_{\cal V}$ as a function of temperature trough eq.~\ref{eq:cv}. First, in the low temperature range, we interpolate the points with the best non-linear curve fit, lowering the chi-squared. Next, we take the temperature numerical derivative and determine the $C_{\cal V}(T_c^-)$.  Second, for the temperatures above $T_c$, we take the data points presenting no measurable condensed fractions and assume them as pure thermal clouds. Under this conditions, the linear behavior of $\Pi$ vs $T$ is well known~\cite{Romero-Rochin2005}, as a result of the dominant Gaussian distribution. We then fit the data linearly and determine the $C_{\cal V}(T_c^+)$, in very good agreement with the data seen in Fig.~\ref{fig:pvst}. In fact, different curves may also fit the data sets, and we have used them to determine the uncertainty presented as error bars in Fig.~\ref{fig:cv}. 

The evolution of the heat capacity as a function of the normalized temperature is plotted in Fig.~\ref{fig:cv}, for two different atom numbers. From there, one may note the sharp change in the measured heat capacity near the critical temperature, $T_{c}$. The lines result from the direct application of equations $10$ and $14$, derived on Ref.~\cite{Grossmann1995}, computing the heat capacity across $T_{c}$. We did use the measured frequencies of our QUIC trap to determine the level spacing, $\hbar \omega / k_{B} \approx 5$nK, and the zero point energy, $E_{0}/k_{B}\approx 10$nK. The absolute values, as well as the steep change in the heat capacity are expected to be number dependent. The measured $C_{\cal V}$ is in very good agreement with the finite-$N$ theory for BECs held in harmonic traps~\cite{Grossmann1995}, as shown in Fig.~\ref{fig:cv}. From that one may conclude that the finite-$N$ corrections, included in the theory~\cite{Grossmann1995}, already contain the essential features shown by our results. 

The choice of testing the results with the non-interacting model is based in the discussions presented by Giorgini~\cite{Giorgini1997}. They concluded that, the inclusion of the 2-body interactions, shall be more pronounced in just a small temperature region, very near $T_{c}$. The end result will be a rounded off curve joining the values just near Tc. It will be very interesting to be able to carefully study the effects of the 2-body interactions in the temperature region ranging from $~0.8$ to $1.0 T/T_{c}$ (see Fig.15 in Ref.~\cite{Giorgini1997}). In doing so, one would be able to map the round off curve joining the two regimes. In any case, we are confident that the method used here is reliable and robust to treat experimental data, regardless of the absolute accuracy achieved. 

\begin{figure}
\centering
 \includegraphics[width=0.7\textwidth]{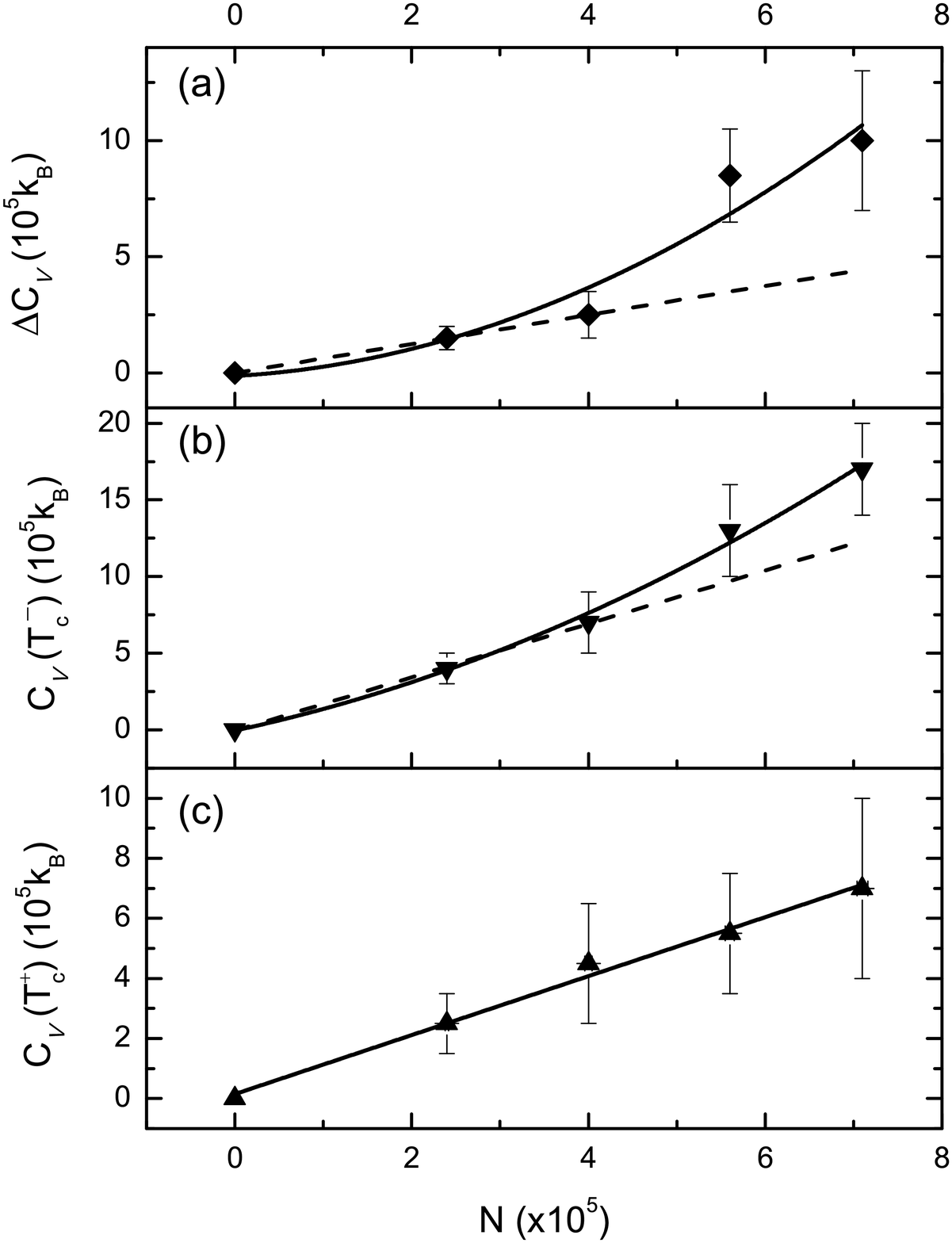}
\caption{
Dependence of the heat capacity around the critical temperature on the total number of atoms: (a) the $\Delta C_{\cal V}$ jump, (b) $C_{\cal V}(T_c^-)$, and (c) $C_{\cal V}(T_c^+)$. Dashed line is the theoretical result for a weakly-interacting, large $N$, case and solid lines are only eyeguides.}
\label{fig:jump}
\end{figure}

The heat capacity evolves, starting from zero, with increasing values proportional to the third power of the normalized temperature, that is: $C^{-}_{\cal V} \propto (T/T_{0})^{3}$, peaking around $0.9$ and $0.98T_{c}$. Very close to $T_{c}$ a steep jump takes place while it goes from $C^{-}_{\cal V}$ to $C^{+}_{\cal V}$ . Right above the critical temperature, a slow decrease with the temperature is observed in $C_{\cal V}$. And, at high temperatures, the heat capacity approaches the temperature independent behavior expected for the non-interacting Bose gas: $3Nk_{B}$. This interesting general shape of the heat capacity is accepted, in the literature~\cite{Ensher1996}, as been characteristic of a second order phase transition. Thus, the investigation of the heat capacity jump of a trapped gas, near $T_{c}$, is important to the understanding of the overall behavior of such phase transition; especially for the non-homogeneous confinement case. We believe, that the opportunity to accurately measure the change in $C_{\cal V}$ very near $T_c$ would allow one to unfold the interactions and the finite number real effects on the transition.

Fig.~\ref{fig:jump}b and Fig.~\ref{fig:jump}c present $C_{\cal V}$ as a function of the trapped atoms number measured across $T_c$. In Fig.~\ref{fig:jump}a, the difference $\Delta C_{\cal V}= C_{\cal V}(T_c^-) - C_{\cal V}(T_c^+)$ is plotted, measured just below/above $T_c$. We observed a small departure from the linear dependence as the number of trapped of atoms increases (Fig.~\ref{fig:jump}a). We believe this might be related to the atomic interactions. On the other hand, the behavior of the heat capacity of a thermal gas (above $T_c$) is quite linear, in very good agreement with the standard theoretical result: $3k_B/N$.

\begin{figure}
\centering
 \includegraphics[width=0.8\textwidth]{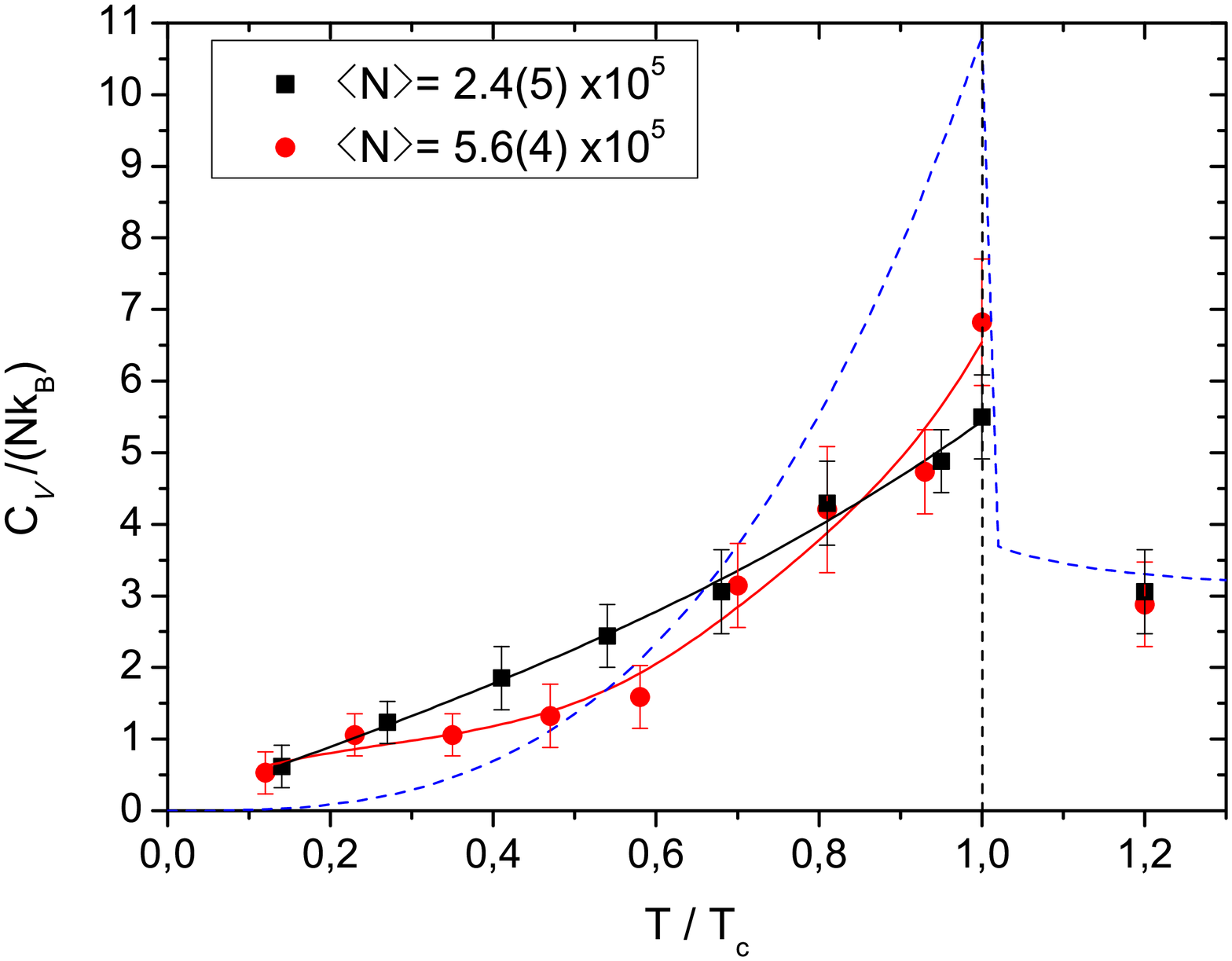}
\caption{
The normalized heat capacity versus the average number of atoms, $\langle N \rangle$, as a function of normalized temperature ($T/T_c$) for two different $\langle N \rangle$. Solid lines are interpolations of the data. The blue dashed line is the theoretical curve for the finite-$N$ corrected, large $N$, case.}
\label{fig:cvnorm}
\end{figure}

In Fig.~\ref{fig:cvnorm}, we plot the normalized heat capacity, $C_{\cal V}/Nk_B$, versus the normalized temperature, $T/T_c$ for two different average number of atoms, $\langle N \rangle$. For an ideal Bose gas, in the limit of large number of atoms, the $C_{\cal V}/Nk_B$ shows the typical, $\langle N \rangle$ independent, curve (dashed blue line). The weakly-interacting BEC data deviates from the universal curve in an intermediate range, below and the critical temperature. We found that $C_{\cal V}/Nk_B$ present larger values for smaller $\langle N \rangle$, in agreement with the theory~\cite{Grossmann1995,Giorgini1997}. The effect of downshifting the $C_{\cal V}$ at large $N$ may seen a bit surprising at first sight but, according to the theoretical models presented on Refs.~\cite{Grossmann1995,Giorgini1997}, it comes from the finite-$N$ correction. The effect is indeed reversed, since the $C_{\cal V}$ correction for smaller $N$ is larger due to the factor proportional to $(T/T_{c})^2\times N^{-1/3}$, see eq.10 in Ref.~\cite{Grossmann1995}. We believe that, close to the zero temperature limit, $T \rightarrow 0$ it will be theoretically allowed to assume an ``energy gap'' behavior~\cite{Huang1987}, which is beyond the scope of this work.

Our method of determining $\Pi$ relies on the ability to indirectly determine the \textit{in situ} number density profile, after some time-of-flight~\cite{Romero-Rochin2012}. The data processing may introduce small shifts in the absolute values, which would be important in a more accurate study of $C_{\cal V}/Nk_B$ at the low temperature limit, as well as near the critical temperature. 
Therefore, perhaps a better methodology for measuring $\Pi$ would be needed to accurately evaluate the behavior of $C_{\cal V}$ in these regions. In this sense, it would be best to be able to take \textit{in situ} images, and to directly determine the 3D density distribution without relying on fittings. Alternative methods to overcome these limitations are currently under discussion, and eventually will allow for more accurate studies.

\section{Conclusions}

We have applied and elaborated an alternative technique for determining the global heat capacity of an inhomogeneous, harmonically trapped $^{87}$Rb Bose gas, not based on the standard LDA. Using the new conjugate (macroscopic) variables \textit{volume parameter} and \textit{pressure parameter}, presented in recent publications~\cite{Romero-Rochin2005,Romero-Rochin2012}, we were able to determine both: the internal energy, and the heat capacity, $C_{\cal V}$, at constant \textit{volume parameter}. We have investigated the evolution of the heat capacity across the transition temperature. A steep $C_{\cal V}$ curve was observed, in the vicinity of $T_{c}$, in close similarity to the $\lambda$-point in liquid $^{4}$He~\cite{Arp2005}. Moreover, the evolution of $C_{\cal V}$ measured near the critical temperature suggests an interplay of the mean field interactions. The remarkable effects are: the $T_{c}$ absolute value downshift; the $C_{\cal V}$ peak round off; and, the larger values of the normalized $C_{\cal V}$ for lower $N$ in a relatively broad T range, below the critical temperature, shown in Fig.~\ref{fig:cvnorm}. Finally, we have briefly discussed the relevance of measuring $C_{\cal V}$ at the low temperature limit, which shall motivate future experiments.

\section{Acknowledgments}

Financial support from FAPESP (program CEPID) as well as from CNPq (program INCT) is appreciated. We thank K.~M.~Magalh\~{a}es and  M.~Caracanhas for technical support.

\bibliography{library}

\end{document}